\documentclass[aps,prd,twocolumn]{revtex4-2}
\usepackage{hyperref}
\usepackage{amsmath}
\usepackage{amssymb}
\usepackage{graphicx}

\begin{document}

\title{\vspace*{10mm}Self-mitigating Trotter circuits for SU(2) lattice gauge theory \\ on a quantum computer}

\author{Sarmed \underline{A Rahman}}
\author{Randy \underline{Lewis}}
\author{Emanuele \underline{Mendicelli}}
\author{Sarah \underline{Powell}}
\affiliation{Department of Physics and Astronomy, York University, Toronto, Ontario, Canada, M3J 1P3 \\}

\date{May 2022. Updated: October 2022.}

\begin{abstract}
\vspace{2mm}
Quantum computers offer the possibility to implement lattice gauge theory in Minkowski rather than Euclidean spacetime, thus allowing calculations of processes that evolve in real time.
In this work, calculations within SU(2) pure gauge theory are able to show the motion of an excitation traveling across a spatial lattice in real time.
This is accomplished by using a simple yet powerful method for error mitigation, where the original circuit is used both forward and backward in time.
For a two-plaquette lattice, meaningful results are obtained from a circuit containing hundreds of CNOT gates.
The same method is used for a five-plaquette lattice, where calculations show that residual systematic effects can be reduced through follow-up mitigation.
\vspace{5mm}

\noindent
DOI of the published version: \href{https://doi.org/10.1103/PhysRevD.106.074502}{10.1103/PhysRevD.106.074502}
\vspace{5mm}

\phantom{.}

\end{abstract}

\maketitle

\section{Introduction}

Lattice gauge theory is a computational framework for obtaining rigorous results from quantum field theories without any recourse to perturbation theory.
A famous example is quantum chromodynamics (QCD), where lattice computations are continuing to reveal novel properties of known hadrons \cite{Aoki:2016frl},
to quantify the prospects for unknown hadrons \cite{Bulava:2022ovd}, and to provide precision input to the search for physics beyond the standard model \cite{Boyle:2022ncb}.

Lattice studies are routinely performed in Euclidean spacetime, meaning that time is imaginary rather than real.
The Euclidean formulation is ideal for a wide range of observables, but a study of dynamics as a function of real time is unfortunately out of scope for this approach.
On the other hand, real-time evolution could be studied quite readily in a Hamiltonian formulation if sufficient computing resources were available,
and this is where future quantum computers are of interest \cite{Bauer:2022hpo}. They have the possibility to store and evolve
quantum states that are much larger than any classical computer could ever attain.
In the present work, we use an IBM quantum computer \cite{ibmq} to calculate the motion of an excitation moving across a lattice in real time.

\begin{figure}
\includegraphics[height=15mm]{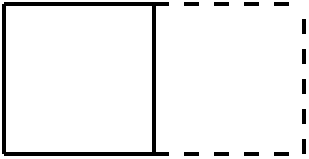}
\caption{In our study, a lattice with 7 gauge links and 2 plaquettes begins at time $t=0$ in the state shown here.
         Each solid line denotes a gauge link with SU(2) irrep $j=\tfrac{1}{2}$.
         Each dashed line denotes a gauge link with SU(2) irrep $j=0$.\label{fig:lattice2}}
\end{figure}

\begin{figure}
\vspace{5mm}
\includegraphics[height=15mm]{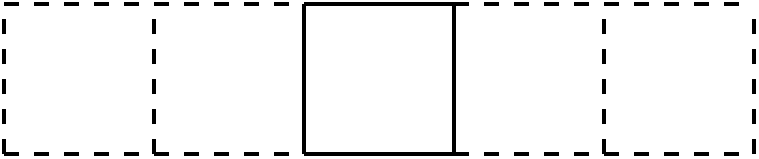}
\caption{In our study, a lattice with 16 gauge links and 5 plaquettes begins at time $t=0$ in the state shown here.
         Each solid line denotes a gauge link with SU(2) irrep $j=\tfrac{1}{2}$.
         Each dashed line denotes a gauge link with SU(2) irrep $j=0$.\label{fig:lattice5}}
\end{figure}

Today's quantum computers have a limited number of qubits, so our study uses a minimal non-Abelian lattice gauge theory.
Specifically, we use an SU(2) gauge group rather than the SU(3) of QCD, and we omit the quarks; thus, the physical particles in the theory are not hadrons but glueballs.
Time is a continuous parameter in the Hamiltonian formulation; however, space is discretized onto a lattice, and the two lattices employed in this work
are shown in Figs.~\ref{fig:lattice2} and \ref{fig:lattice5}.
In general, each gauge link can be in any superposition of the infinite tower of SU(2) irreps $j=0,\tfrac{1}{2},1,\tfrac{3}{2},\ldots$ but, as in other recent quantum computations \cite{Klco:2019evd,ARahman:2021ktn}, the present study truncates to just two basis states: $j=0$ and $j=\tfrac{1}{2}$.
A range of qubit-based approaches to non-Abelian gauge theories can be found in Refs.~\cite{Chandrasekharan:1996ih,Mathur:2004kr,Byrnes:2005qx,Banerjee:2012xg,
Tagliacozzo:2012df,Zohar:2012xf,Zohar:2013zla,Stannigel:2013zka,Anishetty:2014tta,Zohar:2014qma,Mezzacapo:2015bra,Silvi:2016cas,Banuls:2017ena,Banerjee:2017tjn,
Raychowdhury:2018tfj,Sala:2018dui,Raychowdhury:2018osk,Zohar:2019ygc,Raychowdhury:2019iki,Ji:2020kjk,Kasper:2020akk,Davoudi:2020yln,Dasgupta:2020itb,Kasper:2020owz,
Ciavarella:2021nmj,Atas:2021ext,Kan:2021nyu,Zohar:2021nyc,Raychowdhury:2021jbo,Klco:2021lap,Kan:2021xfc,Ciavarella:2021lel,Illa:2022jqb,Gonzalez-Cuadra:2022hxt}.

Notice that we are not using periodic boundary conditions.
Periodicity would include ``round-the-world'' excitations among the list of eigenstates but, at strong coupling where the gauge truncation is most appropriate,
those round-the-world excitations become increasingly negligible
as more plaquettes are added to the lattice.
Also, our calculations are performed on {\tt ibm\_lagos}, which is one of the IBM Quantum Falcon processors \cite{ibmq}, and its qubit layout is best suited to nonperiodic boundaries.

The Hamiltonian for a one-dimensional row of plaquettes is, in the notation of Ref.~\cite{ARahman:2021ktn},
\begin{equation}\label{eq:H}
\hat H = \frac{g^2}{2}\left(\sum_{i={\rm links}}\hat E_i^2 - 2x\sum_{i={\rm plaquettes}}\hat\square_i\right) \,,
\end{equation}
where the only parameter is the gauge coupling $g$, and for convenience we use the notation $x\equiv2/g^4$.
The overall factor of $g^2/2$ can be absorbed into our choice of units for energy.
The first term in $\hat H$ represents the energy stored in the chromoelectric field, and the squared field $\hat E_i^2$ contains an implicit sum
over the three components of the SU(2) Lie algebra.
The second term in $\hat H$ provides the energy stored in the chromomagnetic field (plus a convenient constant) and contains the plaquette
operator $\hat\square_i$ which is the trace of the product of gauge links around the $i$th plaquette.

The initial states chosen for this work are shown in Figs.~\ref{fig:lattice2} and \ref{fig:lattice5}.
The closed loop with $j=\tfrac{1}{2}$ is a gauge-invariant excitation that plays the role of an approximate glueball on these tiny lattices. 
For $x=0$, the chromomagnetic term disappears from the Hamiltonian and our initial states are
eigenstates of the chromoelectric term; thus, the initial states will remain constant in time.
For a small but nonzero value of $x$, the initial excitation will evolve through time to include contributions from other chromoelectric eigenstates and,
because the low-energy eigenstates will dominate, the excitation will travel (with a significant probability) to the next plaquette and so on, across the lattice.
For $x\gtrsim1$, the chromomagnetic term rivals the chromoelectric term meaning the notion of weakly coupled chromoelectric eigenstates is no longer
useful language for time evolution.
For a discussion of similar propagation within an Abelian theory, see Ref.~\cite{Lewis:2019wfx}.
In the present work, we want to observe the non-Abelian phenomena directly through quantum computer calculations for both $x<1$ and $x>1$.

\begin{figure*}
\includegraphics[height=110mm]{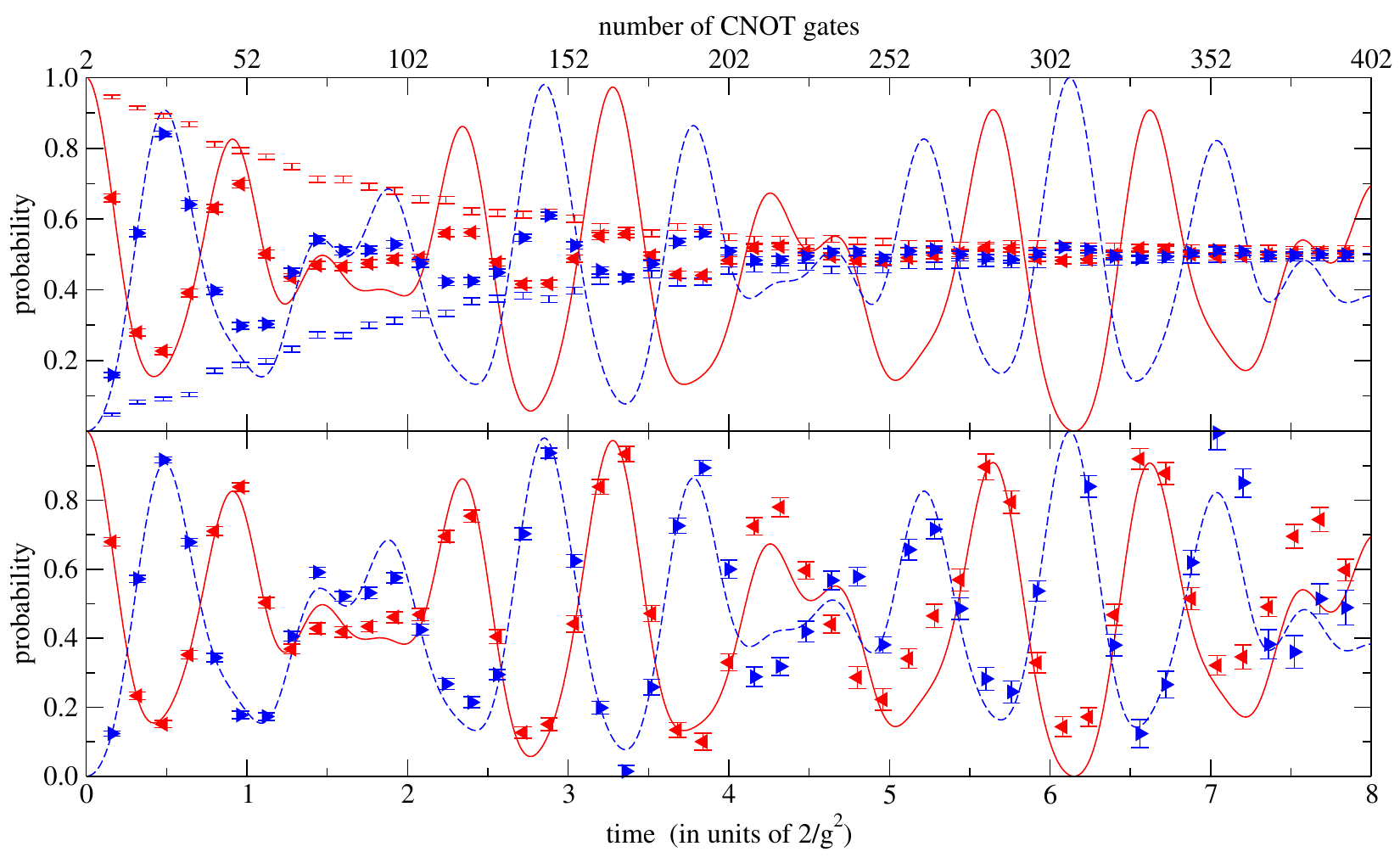}
\caption{Time evolution by self-mitigation on a two-plaquette lattice from the initial state of Fig.~\ref{fig:lattice2} with gauge coupling $x=2.0$
         and time step $dt=0.08$.
         In both panels,
         the red solid (blue dashed) curve is the exact probability of the left (right) plaquette being measured to have $j=\tfrac{1}{2}$.
         {\bf Upper panel:} The red left-pointing (blue right-pointing) triangles are the physics data computed from the {\tt ibm\_lagos} quantum processor.
         The red (blue) error bars without symbols are the mitigation data computed on {\tt ibm\_lagos} from the same circuit but with half the steps forward in time and then half backward in time.
         {\bf Lower panel:} The triangles are the physics results obtained by applying Eq.~(\ref{eq:ratio}) to the data from the upper panel.\label{fig:x20plaq2}}
\end{figure*}

Today's quantum computers are noisy, so significant error rates must be handled.
A modest number of qubits means that full-fledged error correction is not possible, but various options for mitigating errors have been explored by many authors
in several contexts, as reviewed in Ref.~\cite{Endo:2020kro}.
One approach is to use a separate circuit to estimate the errors expected to arise in the actual circuit of
interest \cite{Strikis:2020pcz,Czarnik:2020xic,Lowe,Vovrosh:2021ocf,Urbanek:2021oej,Sopena:2021sai,Rosenberg:2021lpj}.
As described below, the present project follows this approach and is inspired, in particular, by Ref.~\cite{Urbanek:2021oej}.
Our variation on this theme is to use the same circuit both forward and backward in time rather than introducing a separate circuit for error mitigation.

As is well known, a Trotter formula \cite{Lloyd} divides the total time $t$ into small intervals $dt$,
and expresses the time evolution operator $e^{-iHt}$ as an operator product that approximates the evolution with one $dt$ step after another.
Various Trotter formulas can be defined and for each of them the approximation becomes exact as $dt\to0$.
Higher-order Trotter formulas \cite{Hatano:2005gh} will approach the true result with a higher power of $dt$, but the number of gates they require can be prohibitively large.

Figure~\ref{fig:x20plaq2} provides a sequence of calculations performed on an IBM quantum computer.
From left to right in this figure, each data point adds two Trotter steps to the total circuit.
Whereas the pioneering calculation for SU(2) in Ref.~\cite{Klco:2019evd} was limited to one or two Trotter steps in total for a circuit having six CNOT gates, Fig.~\ref{fig:x20plaq2}
demonstrates dozens of Trotter steps totaling a few hundred CNOT gates.
To accomplish this, our computations use a method we call self-mitigation.
Agreement with the exact curves is good but not perfect and, as will be mentioned in the following sections, follow-up mitigation methods can improve these results further.
Nevertheless, self-mitigation by itself has dramatic benefits and is a central component for the present work.

To implement self-mitigation, the desired physics circuit is run twice.
The ``physics run'' applies all $N$ Trotter steps in the forward time direction to arrive at the intended final time where the qubits are measured.
The ``mitigation run'' applies the first $N/2$ Trotter steps forward in time and the remaining $N/2$ steps backward in time ($dt\to-dt$) which would return to the initial state on
error-free hardware.
Measuring the qubits at the end of the mitigation run provides a determination of the errors.
That knowledge can then be used to infer how noise is affecting the physics run, thus allowing the underlying true physics to be extracted with improved accuracy.

In essence, self-mitigation can be viewed as a particular case of the method proposed in Ref.~\cite{Urbanek:2021oej}.
Those authors demonstrated the value of running two circuits: the desired physics circuit and a partner circuit for noise estimation.
Our choice is to use the same circuit in both runs, with opposite signs for $dt$ in the second half.
This means our mitigation computations use the exact same gates in the exact same order as our physics computations;
the only distinction is $dt\to-dt$ in half of the circuit.

Because of the sign change for $dt$ in the second half of the mitigation run, quantum gates receive different input states than in the physics run.
To account for the dependence of CNOT gates on their input states, randomized compiling is used \cite{Wallman}.
This means a randomly selected Pauli or identity gate is applied to each input qubit for the CNOT gate, which is then followed by the particular Pauli/identity pair
that causes the circuit's overall result to be unchanged.
Figure~\ref{fig:x20plaq2} was obtained from 148 physics runs and 148 mitigation runs, each with a different set of CNOT randomizations.

Details of our study can be found in Sec.~\ref{sec:results} for both the two-plaquette and five-plaquette lattices,
and additional information is contained in the appendices.
Brief discussions and an outlook are provided in Sec.~\ref{sec:discussion}.

\section{Results}\label{sec:results}

\subsection{Time evolution on a two-plaquette lattice}

One way to obtain the chromoelectric eigenstates for a row of $N$ plaquettes is to start with $j=0$ at each gauge link
and then apply various sequences of plaquette operators.
Retaining only $j=0$ and $j=\tfrac{1}{2}$ gives $2^N$ basis states in total, and they can be coded into a qubit register by assigning one qubit to each plaquette.
The top and bottom links of the $n$th plaquette are always equal to each other, either both $j=0$ or both $j=\tfrac{1}{2}$.
The qubit encodes that pair of options.
Each vertical link in the lattice is completely specified by its neighboring plaquette values, being $j=0$ if the neighboring
plaquettes are equal to each other and $j=\tfrac{1}{2}$ otherwise.
The two-qubit expression for a two-plaquette lattice that emerges from Eq.~(\ref{eq:H}) is
\begin{eqnarray}
\frac{2}{g^2}H &=& \frac{3}{8}\left(7 - 3Z_0 - Z_0Z_1 - 3Z_1\right) \nonumber \\
                && - \frac{x}{2}(3+Z_1)X_0 - \frac{x}{2}(3+Z_0)X_1 \label{eq:H2orig}
\end{eqnarray}
where $X_n$ and $Z_n$ denote Pauli gates acting on the $n$th qubit.

The time evolution operator is obtained from exponentiation of the Hamiltonian, and any term involving two plaquettes will require entangling gates.
For IBM hardware, the native entangling gate is the CNOT gate which is a controlled Pauli $X$ gate, and because of this we prefer to first express the Hamiltonian in terms of $Y$ and $Z$ gates by applying a $\sqrt{Z}$ rotation to each qubit in the register.
The result is
\begin{eqnarray}\label{eq:H2}
\frac{2}{g^2}H &=& \frac{3}{8}\left(7 - 3Z_0 - Z_0Z_1 - 3Z_1\right) \nonumber \\
                && - \frac{x}{2}(3+Z_1)Y_0 - \frac{x}{2}(3+Z_0)Y_1 \,.
\end{eqnarray}
For this form of the Hamiltonian, just a few basic identities,
\begin{equation}
e^{-i\theta Z_j} = RZ_j(2\theta) \,,
\end{equation}
\begin{equation}
e^{-i\theta Y_j} = RY_j(2\theta) \,,
\end{equation}
\begin{equation}
e^{-i\theta Z_jZ_k} = CX_{jk}RZ_k(2\theta)CX_{jk} \,,
\end{equation}
\begin{equation}
e^{-i\theta Z_jY_k} = CX_{jk}RY_k(2\theta)CX_{jk} \,,
\end{equation}
are sufficient to arrive at the time evolution operator in terms of single-qubit rotations and CNOT gates (here called $CX_{jk}$ for control qubit $j$ and target qubit $k$).
According to Eq.~(\ref{eq:H2}), a first-order Suzuki-Trotter step needs six CNOT gates in general, but this can be reduced to four by ordering the Hamiltonian terms
appropriately.  A second-order Suzuki-Trotter expression can also be constructed with six CNOT gates, and a careful ordering
shrinks that number due to cancellations between neighboring Trotter steps.
See Appendix~\ref{sec:methods} for details.
\begin{figure}[b]
\includegraphics[width=85mm]{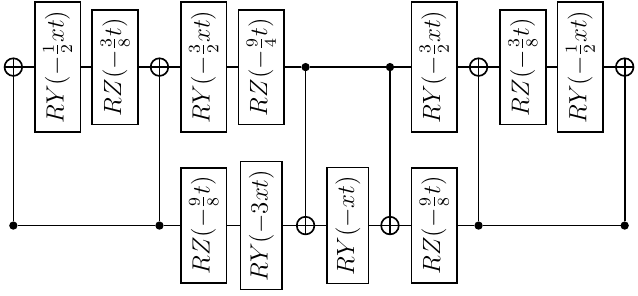}
\caption{Single second-order Suzuki-Trotter step for the two-plaquette lattice.
         The first and last CNOT gates cancel with neighboring Trotter steps, leaving four CNOT gates per Trotter step.
         A forward step has $t=dt$ and a backward step has $t=-dt$.\label{fig:circuit2}}
\end{figure}

We use the second-order Trotter step displayed in Fig.~\ref{fig:circuit2}.
Notice that the CNOT gates at each edge will cancel those from neighboring Trotter steps and are thus not included in our code,
except at the very beginning and very end of the circuit.
After the CNOT pairs cancel, the adjacent $RY(-\tfrac{1}{2}xt)$ gates can be combined into a single $RY(-xt)$.

Due to randomized compiling, each CNOT gate in Fig.~\ref{fig:circuit2} should be replaced with one of the 16 randomized versions listed in Appendix~\ref{sec:randomCNOT}.
The extra Pauli gates introduced by this process can be combined with those from neighboring CNOT gates, and with the rotation gates in between them, to
keep the circuit depth from growing too much.
Appendix \ref{sec:PauliR} provides the explicit relations that we have implemented.  Notice that all of the randomized Pauli gates get absorbed into rotation gates
except a few remaining Pauli $X$ gates.
We chose $X$ as the Pauli gate to retain because it is a native gate on IBM hardware.

Our code submitted 300 runs to the quantum hardware within each job, with $10^4$ hits per run, by sending the 300 circuits to the quantum hardware as a single python list.
This ensures that the mitigation circuits run back to back with the physics circuits, thus experiencing essentially the same hardware conditions.
Four of the runs were for mitigation of measurement errors, 148 were for self-mitigation, and 148 were for the physics calculation.

Measurement mitigation \cite{Bravyi} accounts for errors made during the measurement step at the end of a circuit and
is done in a straightforward manner: Prepare each of the four basis states of the computational basis as a separate circuit, measure each qubit, construct the
4$\times$4 calibration matrix, and apply that matrix to all 296 runs of the physics circuit via sequential least squares programming.
IBM's qiskit software contains a package to handle measurement error mitigation \cite{qiskit} but we chose to write our own equivalent implementation.

The upper panel of Fig.~\ref{fig:x20plaq2} displays the averaged results from the mitigation run and the averaged results from the physics run with gauge coupling $x=2.0$ for the two-plaquette lattice.
The time interval used for each Trotter step is 0.08 in units of $2/g^2$ and symbols on the graph show even numbers of time steps.
For each data point, a statistical error bar for $148$ runs is obtained from 1480 bootstrap samples and then combined in
quadrature with the statistical error from $10^4$ hits per run.
On perfect hardware, the mitigation results would have probability = 1 for the left plaquette and 0 for the right plaquette at all times.
Instead, they smoothly approach the pure noise value, probability = $\tfrac{1}{2}$, as time increases.
The raw physics calculations are always bounded by the mitigation results as expected, since the true physics probabilities are necessarily between 1 and 0.

If the combined effects of self-mitigation and randomization lead to incoherent noise that is independent of the $dt\to-dt$ sign flip, then we can obtain the
true physics result by equating ratios:
\begin{equation}\label{eq:ratio}
\left.\frac{P_{\rm true}-\tfrac{1}{2}}{P_{\rm computed}-\tfrac{1}{2}}\right|_{\rm physics\,run}
=
\left.\frac{P_{\rm true}-\tfrac{1}{2}}{P_{\rm computed}-\tfrac{1}{2}}\right|_{\rm mitigation\,run}.
\end{equation}
\vspace{3mm}

\noindent
For related discussions of depolarizing noise see, for example, Refs.~\cite{Vovrosh:2021ocf,Urbanek:2021oej}.
The application of our ratio to the data plotted in the upper panel of Fig.~\ref{fig:x20plaq2} produces the results displayed in the lower panel.
Comparison of these two plots provides an eye-catching example for the usefulness of self-mitigation.

\subsection{An excitation traveling across a lattice}

Having demonstrated the success of self-mitigation, we now apply it to the physics goal of seeing a traveling excitation.
This is in contrast to Fig.~\ref{fig:x20plaq2} where the large gauge coupling means all eigenvalues made significant contributions and all frequencies were competing in the resulting graph.
To see a traveling excitation, we must use $x\lesssim1$
where the lowest eigenvalues will dominate, meaning that single-plaquette states are of particular importance.
In this case an initial single-plaquette state will move along a row of plaquettes and then reflect from the end of the lattice and propagate back.
The propagation time will be larger for smaller $x$, becoming infinite (so the excitation never moves) at $x=0$.
For an overview of this type of propagation within an Abelian theory, see Ref.~\cite{Lewis:2019wfx}.

\begin{figure}
\includegraphics[height=65mm]{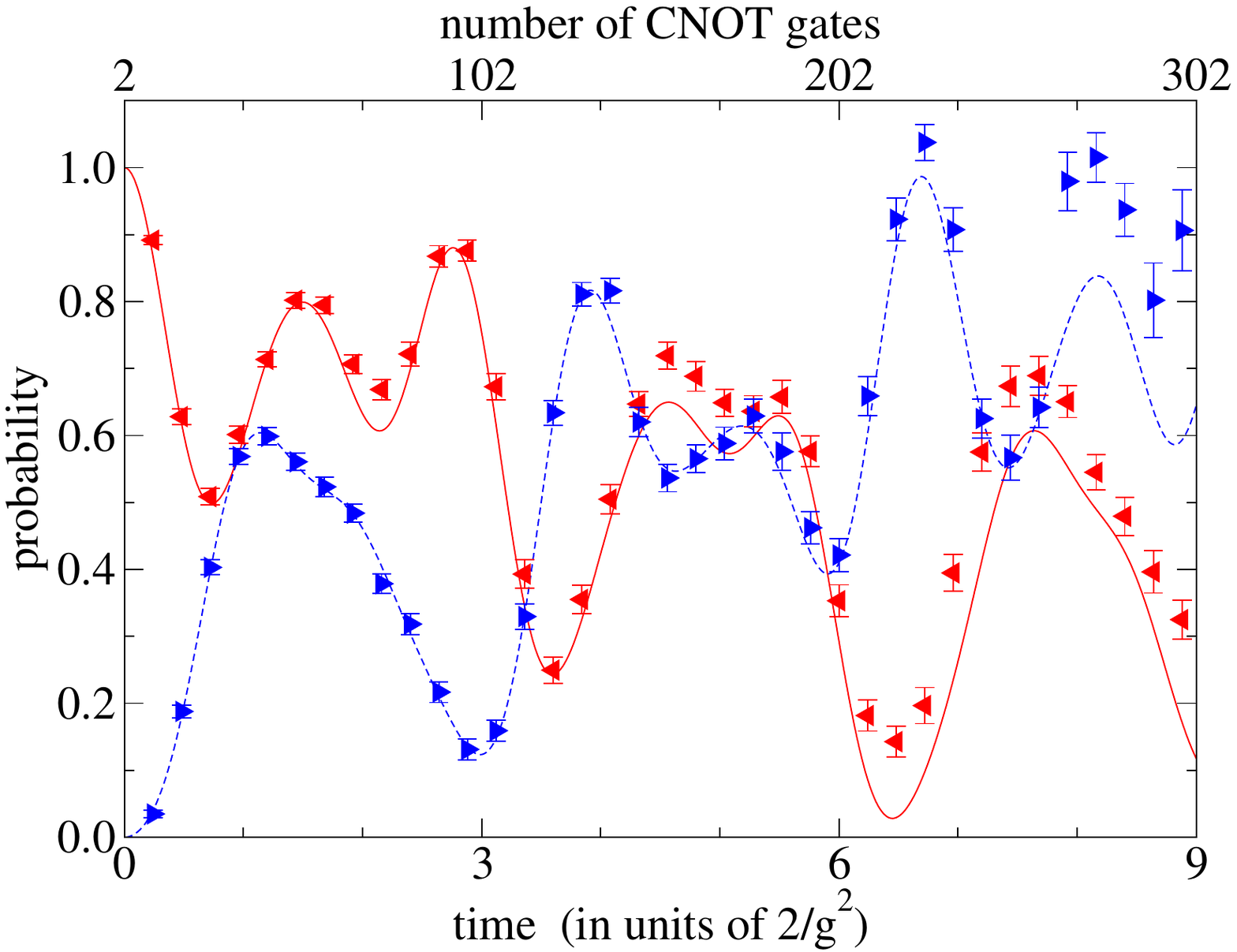}
\caption{Excitation moving from one plaquette to another in real time.
         The initial state of the two-plaquette lattice is Fig.~\ref{fig:lattice2}, gauge coupling is $x=0.8$ and time step is $dt=0.12$.
         The red solid (blue dashed) curve is the exact probability of the left (right) plaquette being measured to have $j=\tfrac{1}{2}$.
         The red left-pointing (blue right-pointing) triangles are the corresponding calculations on the {\tt ibm\_lagos} quantum processor after self-mitigation.\label{fig:x08plaq2}}
\end{figure}

Here we choose $x=0.8$ which is small enough to make the excitation's movement visible and yet large enough to see the movement within a modest time window.
Results are presented in Fig.~\ref{fig:x08plaq2} for a time step of $dt=0.12$.
Higher frequencies are clearly visible as oscillations superimposed on the slow transition of large probability from the left plaquette to the right plaquette.

As in Fig.~\ref{fig:x20plaq2}, each data point in Fig.~\ref{fig:x08plaq2} uses two additional second-order Trotter steps relative to the previous time step.
As the number of CNOT gates approaches 300, the data begin to deviate from the exact curves but the physics conclusion is clear.
The probability of observing an excitation (i.e.\ $j=\tfrac{1}{2}$) at the left plaquette has dropped from 100\% at time $t=0$ to something below 50\%, while
the probability for an excited right plaquette has increased from 0 at time $t=0$ to something above 50\%.

\begin{figure*}
\includegraphics[width=180mm]{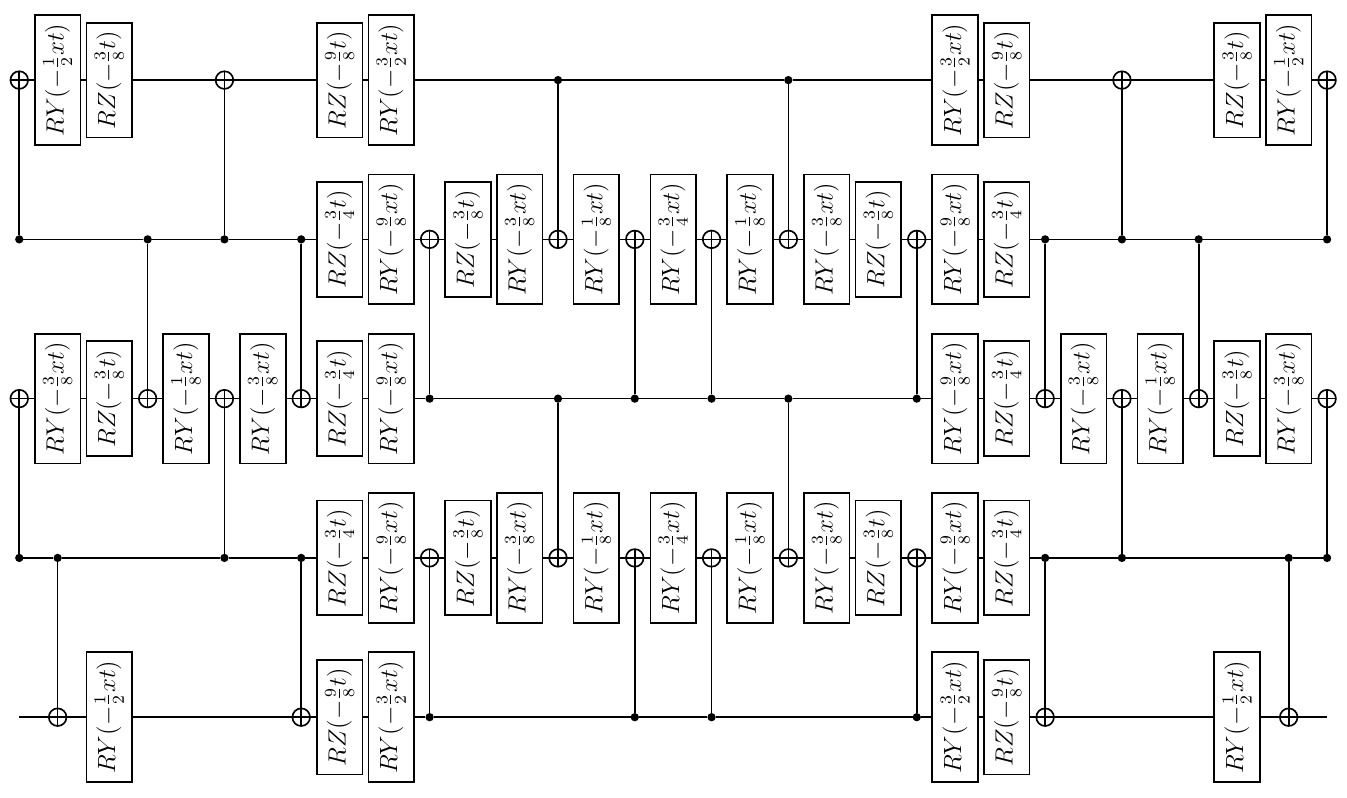}
\caption{Single second-order Suzuki-Trotter step for the five-plaquette lattice.
         The five horizontal lines are the five qubits that represent the five plaquettes in order across the lattice.
         A forward step has $t=dt$ and a backward step has $t=-dt$.\label{fig:circuit5}}
\end{figure*}

\subsection{Time evolution on a five-plaquette lattice}

The {\tt ibm\_lagos} processor has 7 qubits but the longest string of nearest-neighbor couplings joins only 5 qubits, suggesting that we attempt the challenge of a five-plaquette lattice.
Generalizing Eq.~(\ref{eq:H2orig}) to $N$ plaquettes gives
\begin{equation}
H = \frac{g^2}{2}(h_E + h_B) \,,
\end{equation}
\begin{eqnarray}
h_E &=& \frac{3}{8}(3N+1) - \frac{9}{8}(Z_0+Z_{N-1}) - \frac{3}{4}\sum_{n=1}^{N-2}Z_n \nonumber \\
     && - \frac{3}{8}\sum_{n=0}^{N-2}Z_nZ_{n+1} \,,
\end{eqnarray}
\begin{eqnarray}
h_B &=& -\frac{x}{2}(3+Z_1)X_0 - \frac{x}{2}(3+Z_{N-2})X_{N-1} \nonumber \\
     && - \frac{x}{8}\sum_{n=1}^{N-2}(9+3Z_{n-1}+3Z_{n+1}+Z_{n-1}Z_{n+1})X_n \,, \nonumber \\ \label{eq:H5B}
\end{eqnarray}
and here we choose $N=5$.
As described for the two-plaquette case, we will transform the $X$ gates into $Y$ gates for the practical convenience of expressing $H$ with non-$X$ Pauli gates
because the controlled $X$ gate is the native controlled gate for {\tt ibm\_lagos}.

Notice that Eq.~(\ref{eq:H5B}) contains terms that involve three different qubits, arising because application of a plaquette operator requires input from both of the neighboring plaquettes.
This is a new feature for a lattice having more than two plaquettes and is handled with an identity involving four CNOT gates:
\begin{equation}
e^{-i\theta Z_jY_kZ_l} = CX_{lk}CX_{jk}RY_k(2\theta)CX_{jk}CX_{lk} \,.
\end{equation}
A random ordering of the Hamiltonian terms can lead to many CNOT gates within a single Trotter step,
but a particular ordering can reduce the count to 16 CNOT gates in a first-order Trotter step
or 22 CNOT gates in a second-order Trotter step.
We choose the second-order option and our Trotter step is displayed in Fig.~\ref{fig:circuit5}.
Although 28 CNOT gates are shown, only 22 are needed because the three on each end will cancel with those in the neighboring Trotter steps except, of course,
at the very beginning and end of the complete circuit.
After canceling the neighboring CNOT gates, three pairs of $RY$ gates beside them are merged into three single $RY$ gates.

The CNOT gates in Fig.~\ref{fig:circuit5} are placeholders for any randomly chosen options from Appendix~\ref{sec:randomCNOT}.
The extra Pauli gates arising from the randomization are absorbed into rotation gates through the expressions in Appendix~\ref{sec:PauliR}.
Randomized compiling should lead to predominantly incoherent noise, thus allowing the use of self-mitigation.
Four Trotter steps forward in time will produce the physics result.
Two Trotter steps forward followed by two backward will return to the initial state.
The noise is mitigated by putting those two cases into Eq.~(\ref{eq:ratio}).

Our initial state, as shown in Fig.~\ref{fig:lattice5}, is symmetric under an end-to-end reflection of the lattice.
However, our form of the Trotter step does not appear symmetric under that reflection, which corresponds to flipping Fig.~\ref{fig:circuit5} from top to bottom.
This is a consequence of our particular choices for representing and ordering the individual Hamiltonian terms.
The symmetry is maintained up to standard Trotter errors that vanish as $dt$ approaches zero.
The fact that our Trotter step does not comprise a simple repetitive pattern of gates is nice to see because we want to avoid a simplistic special case.
The goal here is to push self-mitigation to its limits on the available quantum hardware.

\begin{figure}
\includegraphics[height=65mm]{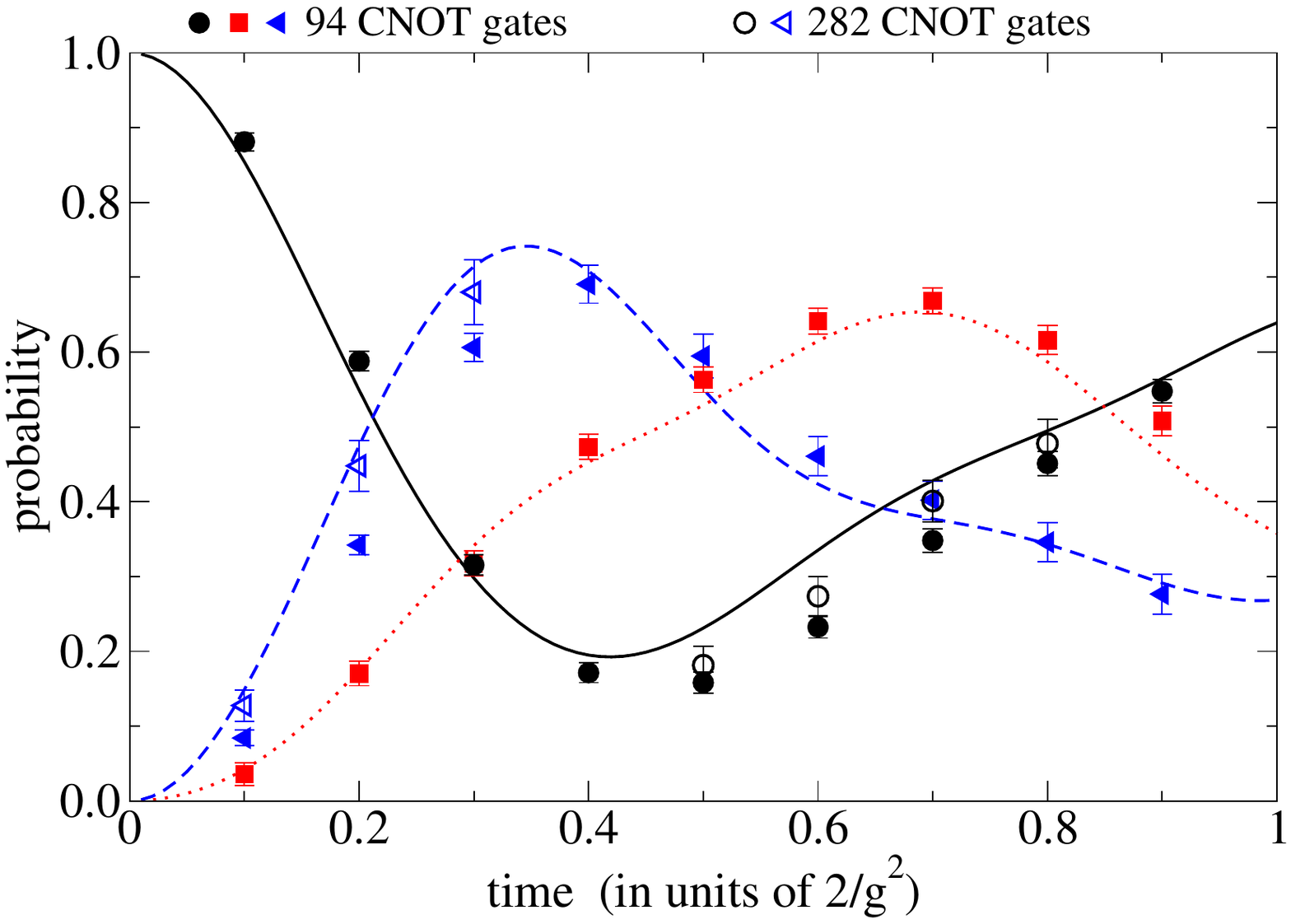}
\caption{Time evolution on a five-plaquette lattice from the initial state of Fig.~\ref{fig:lattice5} with gauge coupling $x=2.0$.
         The black solid, red dotted, and blue dashed curves are respectively the exact probability of measuring $j=\tfrac{1}{2}$ for the center plaquette,
         its neighbor, and the outer plaquette.
         The black circles, red squares, and blue triangles are the corresponding calculations on the {\tt ibm\_lagos} quantum processor after self-mitigation.
         Each filled data point uses four second-order Trotter steps and thus 94 CNOT gates.
         Open data points are augmented by zero-noise extrapolation.\label{fig:x20plaq5}}
\end{figure}

For the two-plaquette lattice we used a constant time step $|dt|$ while changing the number of Trotter steps, but for the five-plaquette lattice
we do the opposite.
Figure~\ref{fig:x20plaq5} shows calculations for various time step sizes, all computed with four of the Trotter steps from Fig.~\ref{fig:circuit5}.
Therefore the total circuit contains 94 randomized CNOT gates.
Each job contains 300 runs (each with $10^4$ hits), where $2^5$=32 runs are used for measurement-error mitigation, 134 runs for the physics calculation, and 134
runs for self-mitigation.
Is this number of runs sufficient?
To address that, four separate jobs were used per data point and analyzed separately.
Because the variations among them were comparable to the statistical error bars, the four results were averaged to produce Fig.~\ref{fig:x20plaq5}.

Most of the solid symbols in Fig.~\ref{fig:x20plaq5} are consistent with the true curves but some systematic deviations are apparent, particularly for
the center plaquette at times between 0.5 and 0.8 and the outer plaquette at times between 0.1 and 0.3.
These residual effects can be handled by using additional error mitigation beyond self-mitigation, specifically zero-noise extrapolation \cite{Li}, as shown by the open symbols in Fig.~\ref{fig:x20plaq5}.
To obtain the open symbols, each CNOT gate was replaced by a triplet of identical CNOT gates and that circuit was studied with self-mitigation and measurement error
mitigation in our standard fashion.
A linear extrapolation of those results (triplet CNOT noise) and the solid symbols (singlet CNOT noise) to the target point of zero CNOT noise produced
the open symbols.
It is encouraging to see that they lie closer to the true curves.
A higher-order extrapolation would allow for a more detailed study of these residual effects but this is beyond the aims of the current project.
Notice that the triplet-CNOT circuit already contains 282 CNOT gates and therefore represents another success for self-mitigation.

Because this calculation uses a gauge coupling $x=2.0$, the results are not interpreted as a traveling excitation.
Indeed, the three peaks in Fig.~\ref{fig:x20plaq5} are not ordered from the starting location (i.e.\ the center of the five-plaquette lattice as shown in Fig.~\ref{fig:lattice5})
out to the ends of the lattice.
Instead, $x=2.0$ means chromoelectric and chromomagnetic contributions are both important and all frequencies play a role.
Self-mitigation has provided a useful reflection of the true physics, while the remaining systematic errors can be managed through zero-noise extrapolation.

In principle, we could now select $x<1$ and look for a traveling excitation on the five-plaquette lattice.
However, this would require significantly more Trotter steps than were used in Fig.~\ref{fig:x20plaq5} and we expect it to
be beyond the reach of present computing hardware even with self-mitigation and zero-noise extrapolation.
Rapid hardware progress will undoubtedly continue, underscoring the importance of having mitigation methods tested and ready.

\section{Discussion}\label{sec:discussion}

This work has used a quantum computer to study time evolution in a non-Abelian lattice gauge theory,
and presents the first observation of a local excitation moving across a spatial lattice.
It also shows time evolution further from the strong coupling regime where a larger number of eigenvalues can play a significant role.
The results span a much larger time range than has been attained from previous qubit calculations in non-Abelian gauge theories \cite{Klco:2019evd,Ciavarella:2021nmj,ARahman:2021ktn,Illa:2022jqb}.

The improvement was made possible by a recent major advance in the mitigation of gate errors, particularly the idea of defining a noise-mitigation circuit
that resembles the target circuit of physical interest \cite{Urbanek:2021oej}.
Our approach, referred to as self-mitigation, uses the physics circuit as its own noise-mitigation circuit.
This means our mitigation circuit is as close to the desired physics circuit as possible because all gates are retained and kept in exactly the same locations within each run of the circuit.
In a ``physics run,'' all Trotter steps evolve forward in time and measurements are made at the desired final time.
In a ``mitigation run'', half of the Trotter steps evolve forward in time and half evolve backward; thus, measurements are made after returning to the initial time.
Because the initial conditions are known, the mitigation run determines the effects of noise.
That noise is used to mitigate the physics run, thus revealing the physics signal.
For additional information, see Appendix~\ref{sec:methods}.

A well-known alternative, called zero-noise extrapolation \cite{Li}, purposely introduces various amounts of additional noise into the circuit and then
extrapolates to the zero-noise limit.
This additional noise typically reduces the number of Trotter steps that can be handled.
In contrast, self-mitigation has the notable advantage of not introducing any additional noise.
Still, if extra mitigation is desired then zero-noise extrapolation can be applied after self-mitigation, as demonstrated by Fig.~\ref{fig:x20plaq5}.

Randomized compiling is an integral part of self-mitigation, so gate errors can be converted into incoherent errors instead of retaining systematic dependences on
the qubit states being input to each CNOT gate.
The randomization is done through Pauli gates and the conversion to incoherent errors is not exact but is nevertheless of great practical value.

Some variations in the implementation of self-mitigation can be considered.
For example, imagine choosing mitigation runs that have one step forward in time followed by one step backward in time, alternating forward and backward steps until the end of the circuit.
This means the qubit register will never stray far from its initial state but it quickly deviates from the physics run.
Since our goal is to keep the mitigation runs as similar as possible to the physics runs, this is not a helpful scenario.

Another possible variation is to use mitigation runs that have twice as many Trotter steps as the physics runs.
For a physics circuit having $N$ Trotter steps, imagine a mitigation circuit comprising $N$ Trotter steps forward in time followed by $N$ steps backward.
This has the advantage that the first half of the mitigation circuit is identical to the entire physics circuit.
In a sense, the second half is identical to the conjugate of the physics circuit but each CNOT gate has a different input state when moving toward the initial state rather than
away from it, leaving no clear advantage over the implementation used in our study.
Moreover, the mitigation circuit now contains more gates which will reduce the number of Trotter steps that can be attained.
Equation~(\ref{eq:ratio}) could be modified to handle the longer mitigation circuit, for example, by taking the square root of the right-hand side to reflect
a standard assumption about the scaling of errors.
No such assumption is required for the self-mitigation that has been implemented in our work.

\begin{acknowledgments}
This work was supported in part by a Discovery Grant from the Natural Sciences and Engineering Research Council (NSERC) of Canada
and by a NSERC Undergraduate Student Research Award.
We acknowledge the use of IBM Quantum services for this work. The views expressed are those of the authors, and do not reflect the official policy or position of IBM or the IBM Quantum team.
\end{acknowledgments}

\appendix

\section{METHODS}\label{sec:methods}

This section provides extra details about building the circuit, writing the code, running the code, interpreting the results,
and assessing the impact of gauge truncation.
For this discussion, the two-plaquette lattice is used as the explicit example.

{\bf Building the circuit:}
There are three terms in Eq.~(\ref{eq:H2}) that involve two qubits.
When computing time evolution, those terms require CNOT gates.
In our codes, we put those terms side by side at the ends of the list so neighboring CNOT gates can cancel with each other
and with neighboring Trotter steps. This leads to
\begin{eqnarray}
\frac{2}{g^2}H &=& -\frac{x}{2}Z_1Y_0 - \frac{3}{8}Z_0Z_1 - \frac{3}{2}xY_0 - \frac{9}{8}Z_1 \nonumber \\
               & & - \frac{9}{8}Z_0 - \frac{3}{2}xY_1 - \frac{x}{2}Z_0Y_1 \,.
\end{eqnarray}
In a second-order Trotter step, the exponentials of these terms appear twice (each instance with half
of the coefficient), resulting in
\begin{eqnarray}
e^{-iHt} &=& e^{i(xt/4)Z_1Y_0}e^{i(3t/16)Z_0Z_1}e^{i(3xt/4)Y_0}e^{i(9t/16)Z_1} \nonumber \\
         & & e^{i(9t/16)Z_0}e^{i(3xt/4)Y_1}e^{i(xt/4)Z_0Y_1}e^{i(xt/4)Z_0Y_1} \nonumber \\
         & & e^{i(3xt/4)Y_1}e^{i(9t/16)Z_0}e^{i(9t/16)Z_1}e^{i(3xt/4)Y_0} \nonumber \\
         & & e^{i(3t/16)Z_0Z_1}e^{i(xt/4)Z_1Y_0} \,,
\end{eqnarray}
where time $t$ is in units of $2/g^2$.
This result corresponds to the circuit displayed in Fig.~\ref{fig:circuit2}.
Some gates at the center of the product commute with each other, so those have been rearranged and combined in the figure.

{\bf Writing the code:}
A code for self-mitigation can be written with standard qiskit commands.
A sample code for calculating self-mitigated time evolution on a two-plaquette lattice is provided in Ref.~\cite{duet}.

{\bf Running the code:}
The amount of time needed for computations on the qubit register depends on the number of circuits,
the number of shots per circuit, and the number of Trotter steps in each circuit.
Each time value in Fig.~\ref{fig:x20plaq2} comes from one job having 300 circuits with $10^4$ shots per circuit.
For two examples, we note that the job with 4 Trotter steps used 13.7 minutes on the qubit register, and the
job with 40 Trotter steps used 17.0 minutes.
For comparison, a job with four Trotter steps of our five-plaquette computation (again with 300 circuits and $10^4$ shots
per circuit) used 14.5 minutes.

{\bf Interpreting the results:}
A quick glance at the lower panel of Fig.~\ref{fig:x20plaq2} might give the impression that self-mitigation is doing less well 
for probabilities near $\tfrac{1}{2}$, but this is not the case.
\begin{figure}
\includegraphics[height=65mm]{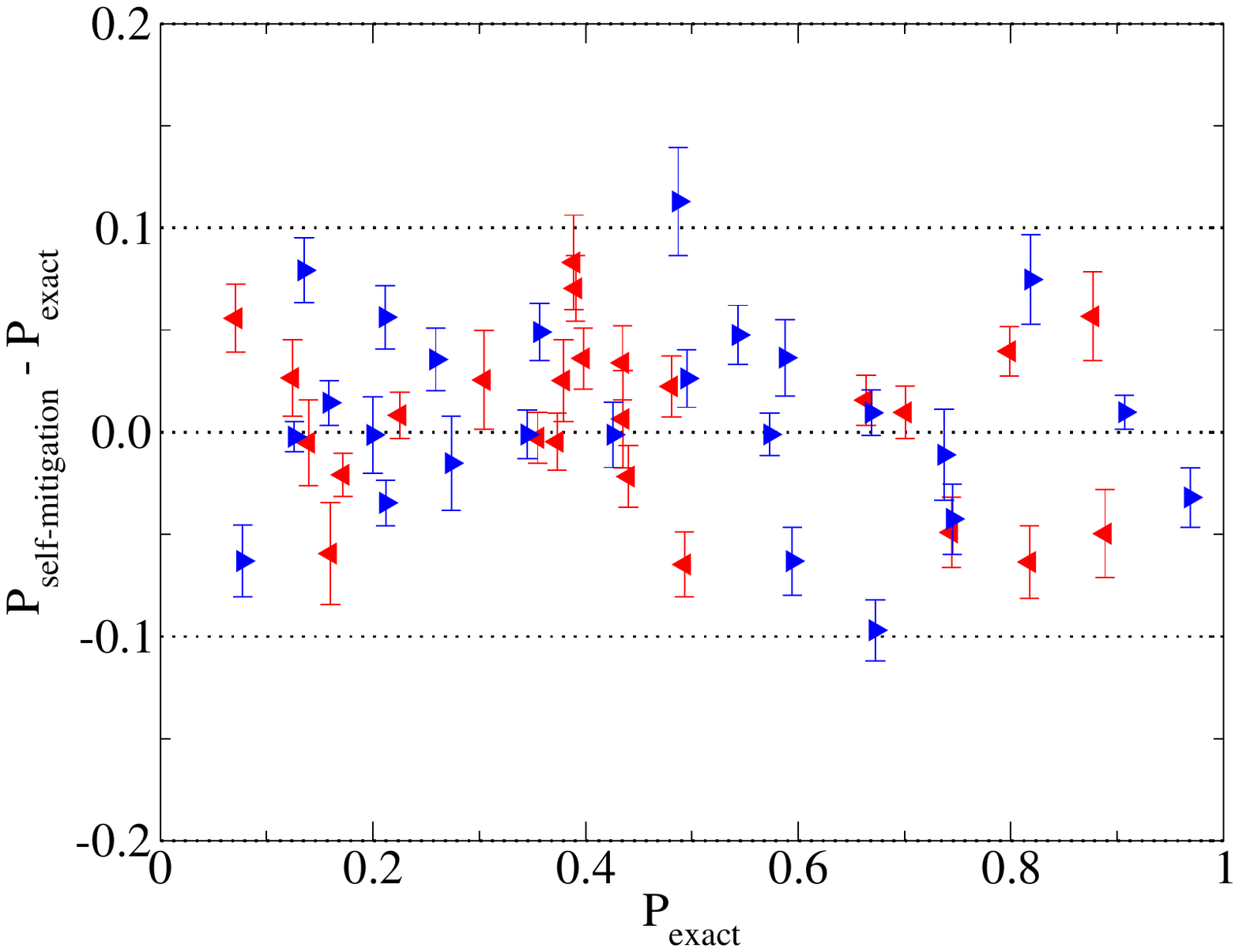}
\caption{All data for $t\leq4$ from the lower panel of Fig.~\ref{fig:x20plaq2} displayed as the difference between
         the self-mitigated computation and the exact probability.  All data points lie within a horizontal band on the graph,
         without any conspicuous dependence on $P_{\rm exact}$, suggesting that self-mitigation
         is performing equally well for all values of the exact probability.\label{fig:discrepancy}}
\end{figure}
\begin{figure}[b]
\includegraphics[height=65mm]{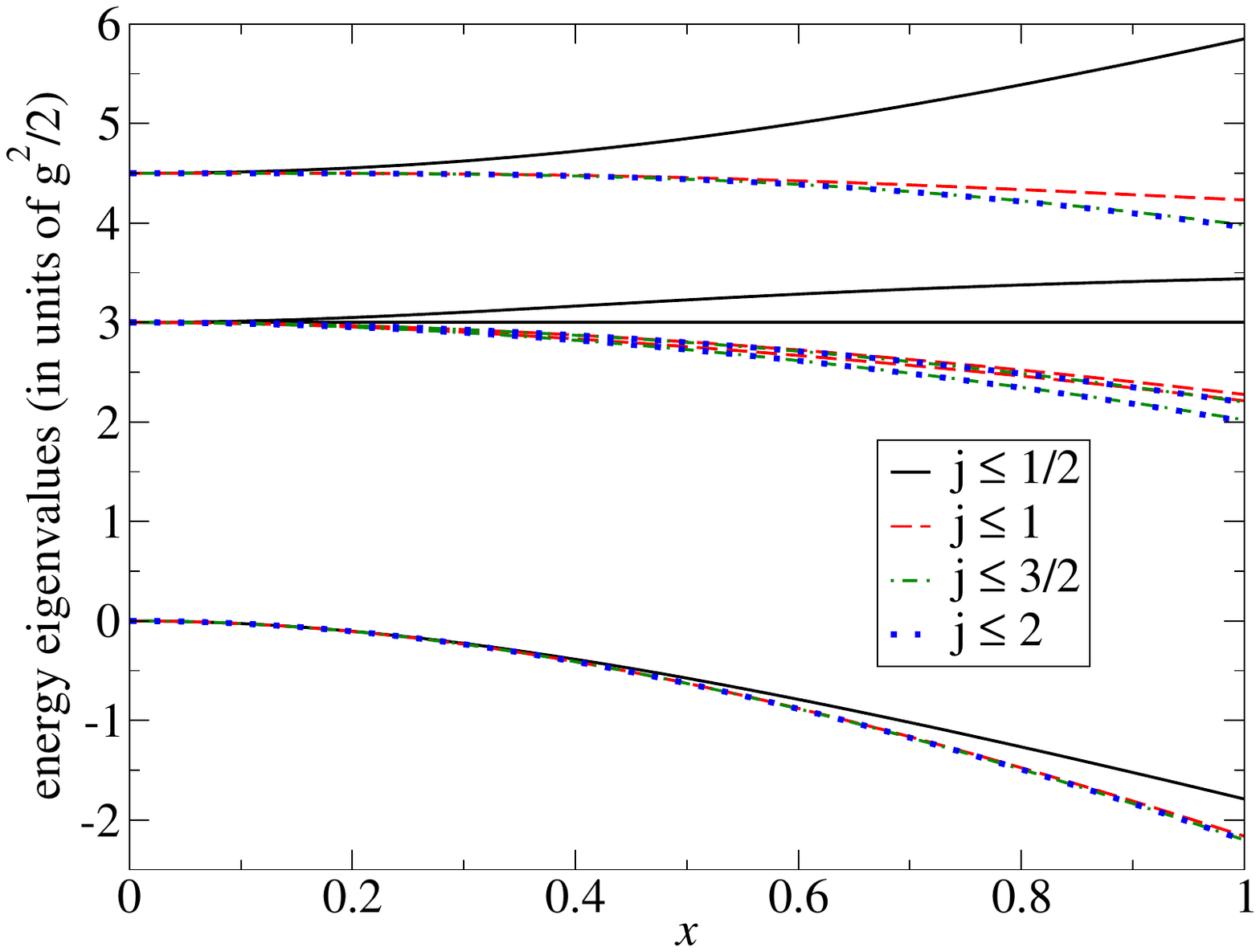}
\caption{The four lowest eigenvalues of the Hamiltonian matrix for a two-plaquette lattice are calculated with various choices
         for the gauge truncation.  Rapid convergence is observed for these $x$ values.
         This graph was obtained from a purely classical computation.\label{fig:eigenvalues}}
\end{figure}
For added clarity, we can replot the data from the left half of that lower panel using different axes.
Figure~\ref{fig:discrepancy} shows the discrepancy between the computed data points and the exact curve as a function of the exact curve's
value.  The graph confirms that self-mitigation is equivalently successful across the full range of probability values.
The same conclusion is true at larger times (not shown in Fig.~\ref{fig:discrepancy}) even though self-mitigating computations at all probabilities begin to deviate further from
the exact probability.

\begin{figure}
\includegraphics[height=65mm]{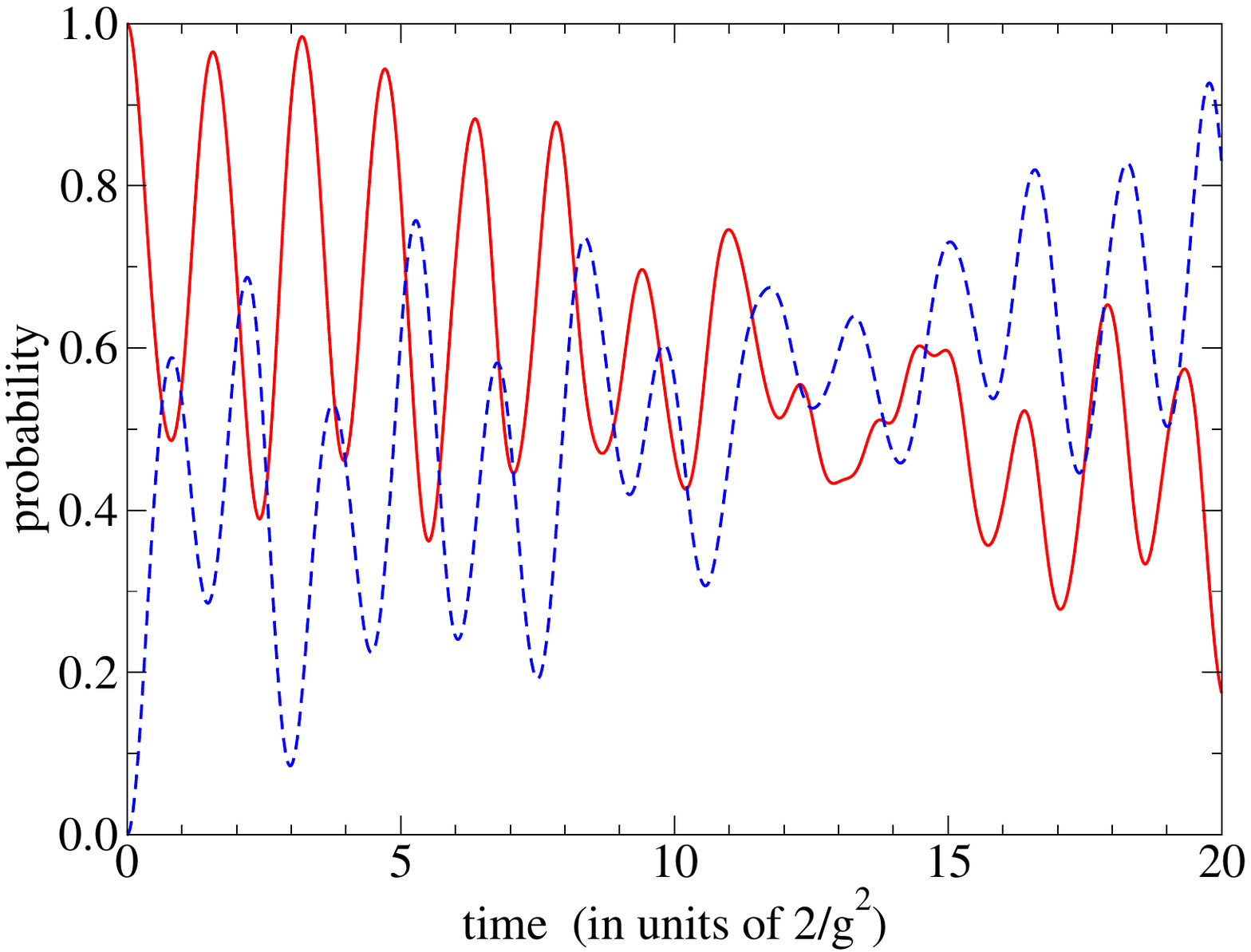}
\caption{An excitation moving from one plaquette to another in real time on a two-plaquette lattice,
         retaining all states having $j\leq2$.  The gauge coupling is $x=0.8$.
         The red solid (blue dashed) curve is the probability of the leftmost (rightmost) gauge link being measured to have $j>0$.
         This graph was obtained from a purely classical computation.\label{fig:jequals2}}
\end{figure}

{\bf Impact of gauge truncation:}
The methods described above allow the observation of an excitation moving across the lattice, as shown for gauge coupling
$x=0.8$ in Fig.~\ref{fig:x08plaq2}.
These results were obtained by truncating the Hilbert space such that each gauge link has only two basis states: $j=0$ and
$j=\tfrac{1}{2}$.  To assess the impact of this truncation, Fig.~\ref{fig:eigenvalues} shows the lowest energy eigenvalues as
the maximum $j$ is increased.  Convergence is rapid at $x=0.8$, with precise results appearing already by $j=2$.
How would the traveling excitation of Fig.~\ref{fig:x08plaq2} be changed if these extra $j$ options had been retained?
As shown in Fig.~\ref{fig:jequals2}, the physics phenomenon is still observed but the transition from left excitation to
right excitation occurs at a different numerical value of the time.

\section{RANDOMIZED CNOT GATES}\label{sec:randomCNOT}

Randomized compiling means that each CNOT gate shown in Fig.~\ref{fig:circuit2} gets replaced by one of the following 16 options, chosen randomly:
\begin{align}
&CX_{jk} \,, \\
& \nonumber \\
X_k &CX_{jk} X_k \,, \\
& \nonumber \\
Y_k &CX_{jk} Z_j Y_k \,, \\
& \nonumber \\
Z_k &CX_{jk} Z_j Z_k \,,
\end{align}
\begin{align}
X_j &CX_{jk} X_j X_k \,, \\
& \nonumber \\
X_j X_k &CX_{jk} X_j \,, \\
& \nonumber \\
X_j Y_k &CX_{jk} Y_j Z_k \,, \\
& \nonumber \\
X_j Z_k &CX_{jk} Y_j Y_k \,, \\
& \nonumber \\
Y_j &CX_{jk} Y_j X_k \,, \\
& \nonumber \\
Y_j X_k &CX_{jk} Y_j \,, \\
& \nonumber \\
Y_j Y_k &CX_{jk} X_j Z_k \,, \\
& \nonumber \\
Y_j Z_k &CX_{jk} X_j Y_k \,, \\
& \nonumber \\
Z_j &CX_{jk} Z_j \,, \\
& \nonumber \\
Z_j X_k &CX_{jk} Z_j X_k \,, \\
& \nonumber \\
Z_j Y_k &CX_{jk} Y_k \,, \\
& \nonumber \\
Z_j Z_k &CX_{jk} Z_k \,,
\end{align}
where $j$ is the control qubit and $k$ is the target qubit.
\begin{table}[b]
\caption{The complete set of options used in Eq.~(\ref{eq:Y}).}\label{tab:Y}
\begin{ruledtabular}
\begin{tabular}{cccc}
$P_1$ & $P_2$ & $\theta$ & $P_3$ \\
\hline
$1$ & $1$ & $\alpha$     & $1$ \\
$1$ & $X$ & $\alpha$     & $X$ \\
$1$ & $Y$ & $\pi+\alpha$ & $1$ \\
$1$ & $Z$ & $\pi+\alpha$ & $X$ \\
$X$ & $1$ & $-\alpha$    & $X$ \\
$X$ & $X$ & $-\alpha$    & $1$ \\
$X$ & $Y$ & $\pi-\alpha$ & $X$ \\
$X$ & $Z$ & $\pi-\alpha$ & $1$ \\
$Y$ & $1$ & $\pi+\alpha$ & $1$ \\
$Y$ & $X$ & $\pi+\alpha$ & $X$ \\
$Y$ & $Y$ & $\alpha$     & $1$ \\
$Y$ & $Z$ & $\alpha$     & $X$ \\
$Z$ & $1$ & $\pi-\alpha$ & $X$ \\
$Z$ & $X$ & $\pi-\alpha$ & $1$ \\
$Z$ & $Y$ & $-\alpha$    & $X$ \\
$Z$ & $Z$ & $-\alpha$    & $1$
\end{tabular}
\end{ruledtabular}
\end{table}

\section{ABSORBING PAULI GATES INTO ROTATIONS}\label{sec:PauliR}

\begin{table}[t]
\caption{The complete set of options used in Eq.~(\ref{eq:Z}).}\label{tab:Z}
\begin{ruledtabular}
\begin{tabular}{cccc}
$P_1$ & $P_2$ & $\theta$ & $P_3$ \\
\hline
$1$ & $1$ & $\alpha$     & $1$ \\
$1$ & $X$ & $\alpha$     & $X$ \\
$1$ & $Y$ & $\pi+\alpha$ & $X$ \\
$1$ & $Z$ & $\pi+\alpha$ & $1$ \\
$X$ & $1$ & $-\alpha$    & $X$ \\
$X$ & $X$ & $-\alpha$    & $1$ \\
$X$ & $Y$ & $\pi-\alpha$ & $1$ \\
$X$ & $Z$ & $\pi-\alpha$ & $X$ \\
$Y$ & $1$ & $\pi-\alpha$ & $X$ \\
$Y$ & $X$ & $\pi-\alpha$ & $1$ \\
$Y$ & $Y$ & $-\alpha$    & $1$ \\
$Y$ & $Z$ & $-\alpha$    & $X$ \\
$Z$ & $1$ & $\pi+\alpha$ & $1$ \\
$Z$ & $X$ & $\pi+\alpha$ & $X$ \\
$Z$ & $Y$ & $\alpha$     & $X$ \\
$Z$ & $Z$ & $\alpha$     & $1$
\end{tabular}
\end{ruledtabular}
\end{table}
\begin{table}[b]
\caption{The complete set of options used in Eqs.~(\ref{eq:YZ}) and (\ref{eq:ZY}).}\label{tab:YZ}
\begin{ruledtabular}
\begin{tabular}{cccc}
$P_1$ & $P_2$ & $\theta_1$ & $\theta_2$ \\
\hline
$1$ & $1$ & $\alpha$     & $\beta$ \\
$1$ & $X$ & $\pi+\alpha$ & $\pi-\beta$ \\
$1$ & $Y$ & $\pi+\alpha$ & $-\beta$ \\
$1$ & $Z$ & $\alpha$     & $\pi+\beta$ \\
$X$ & $1$ & $\pi-\alpha$ & $\pi+\beta$ \\
$X$ & $X$ & $-\alpha$    & $-\beta$ \\
$X$ & $Y$ & $-\alpha$    & $\pi-\beta$ \\
$X$ & $Z$ & $\pi-\alpha$ & $\beta$ \\
$Y$ & $1$ & $\pi+\alpha$ & $\beta$ \\
$Y$ & $X$ & $\alpha$     & $\pi-\beta$ \\
$Y$ & $Y$ & $\alpha$     & $-\beta$ \\
$Y$ & $Z$ & $\pi+\alpha$ & $\pi+\beta$ \\
$Z$ & $1$ & $-\alpha$    & $\pi+\beta$ \\
$Z$ & $X$ & $\pi-\alpha$ & $-\beta$ \\
$Z$ & $Y$ & $\pi-\alpha$ & $\pi-\beta$ \\
$Z$ & $Z$ & $-\alpha$    & $\beta$
\end{tabular}
\end{ruledtabular}
\end{table}
\begin{table}[t]
\caption{The complete set of options used in Eq.~(\ref{eq:ZYZ}).}\label{tab:ZYZ}
\begin{ruledtabular}
\begin{tabular}{ccccc}
$P_1$ & $P_2$ & $\theta_1$ & $\theta_2$ & $\theta_3$ \\
\hline
$1$ & $1$ & $\alpha$     & $\beta$     & $\alpha$ \\
$1$ & $X$ & $\alpha$     & $\pi+\beta$ & $\pi-\alpha$ \\
$1$ & $Y$ & $\alpha$     & $\pi+\beta$ & $-\alpha$ \\
$1$ & $Z$ & $\alpha$     & $\beta$     & $\pi+\alpha$ \\
$X$ & $1$ & $\pi-\alpha$ & $\pi+\beta$ & $\alpha$ \\
$X$ & $X$ & $-\alpha$    & $-\beta$    & $-\alpha$ \\
$X$ & $Y$ & $-\alpha$    & $-\beta$    & $\pi-\alpha$ \\
$X$ & $Z$ & $-\alpha$    & $\pi-\beta$ & $\alpha$ \\
$Y$ & $1$ & $-\alpha$    & $\pi+\beta$ & $\alpha$ \\
$Y$ & $X$ & $\pi-\alpha$ & $-\beta$    & $-\alpha$ \\
$Y$ & $Y$ & $-\alpha$    & $\beta$     & $-\alpha$ \\
$Y$ & $Z$ & $-\alpha$    & $\pi+\beta$ & $\pi+\alpha$ \\
$Z$ & $1$ & $\pi+\alpha$ & $\beta$     & $\alpha$ \\
$Z$ & $X$ & $\alpha$     & $\pi-\beta$ & $-\alpha$ \\
$Z$ & $Y$ & $\pi+\alpha$ & $\pi+\beta$ & $-\alpha$ \\
$Z$ & $Z$ & $\alpha$     & $-\beta$    & $\alpha$
\end{tabular}
\end{ruledtabular}
\end{table}
To absorb neighboring Pauli gates into a rotation around the $y$ axis, retaining the IBM native gate $X$ where required, we make the replacement
\begin{equation}\label{eq:Y}
P_1\,RY(\alpha)\,P_2 \to RY(\theta)\,P_3 \,,
\end{equation}
where the $P_j$ and $\theta$ are given in Table~\ref{tab:Y}.
Each of these replacements is an equality up to an unphysical overall phase.
The corresponding expression for rotation around the $z$ axis is
\begin{equation}\label{eq:Z}
P_1\,RZ(\alpha)\,P_2 \to RZ(\theta)\,P_3 \,,
\end{equation}
where the $P_j$ and $\theta$ are given in Table~\ref{tab:Z}.
The expressions for rotation around both $y$ and $z$ are
\begin{equation}
P_1\,RY(\alpha)RZ(\beta)\,P_2 \to RY(\theta_1)RZ(\theta_2) \,, \label{eq:YZ}
\end{equation}
\begin{equation}
P_2\,RZ(\beta)RY(\alpha)\,P_1 \to RZ(\theta_2)RY(\theta_1) \,, \label{eq:ZY}
\end{equation}
where the $P_j$ and $\theta_j$ are given in Table~\ref{tab:YZ}.
The expressions for rotation around $z$ then $y$ then $z$ are
\begin{equation}\label{eq:ZYZ}
P_1\,RZ(\alpha)RY(\beta)RZ(\alpha)\,P_2 \to RZ(\theta_1)RY(\theta_2)RZ(\theta_3) \,,
\end{equation}
where the $P_j$ and $\theta_j$ are given in Table~\ref{tab:ZYZ}.


\begin{thebibliography}{99}

\bibitem{Aoki:2016frl}
S.~Aoki, Y.~Aoki, D.~Becirevic, C.~Bernard, T.~Blum, G.~Colangelo, M.~Della Morte, P.~Dimopoulos, S.~D\"urr and H.~Fukaya, \textit{et al.},
Review of lattice results concerning low-energy particle physics,
\href{https://doi.org/10.1140/epjc/s10052-016-4509-7}{Eur. Phys. J. C \textbf{77}, 112 (2017)}.
%[arXiv:1607.00299 [hep-lat]].

\bibitem{Bulava:2022ovd}
J.~Bulava, R.~Brice\~no, W.~Detmold, M.~D\"oring, R.~G.~Edwards, A.~Francis, F.~Knechtli, R.~Lewis, S.~Prelovsek and S.~M.~Ryan, \textit{et al.},
Hadron Spectroscopy with Lattice QCD,
\href{https://arXiv.org/abs/2203.03230}{arXiv:2203.03230}.

\bibitem{Boyle:2022ncb}
P.~Boyle, D.~Bollweg, R.~Brower, N.~Christ, C.~DeTar, R.~Edwards, S.~Gottlieb, T.~Izubuchi, B.~Joo and F.~Joswig, \textit{et al.},
Lattice QCD and the Computational Frontier,
\href{https://arXiv.org/abs/2204.00039}{arXiv:2204.00039}.

\bibitem{Bauer:2022hpo}
C.~W.~Bauer, Z.~Davoudi, A.~B.~Balantekin, T.~Bhattacharya, M.~Carena, W.~A.~de Jong, P.~Draper, A.~El-Khadra, N.~Gemelke and M.~Hanada, \textit{et al.},
Quantum Simulation for High Energy Physics,
\href{https://arXiv.org/abs/2204.03381}{arXiv:2204.03381}.

\bibitem{ibmq} IBM Quantum, \href{https://quantum-computing.ibm.com}{https://quantum-computing.ibm.com}.

\bibitem{Klco:2019evd}
N.~Klco, J.~R.~Stryker and M.~J.~Savage,
SU(2) non-Abelian gauge field theory in one dimension on digital quantum computers,
\href{https://doi.org/10.1103/PhysRevD.101.074512}{Phys. Rev. D \textbf{101}, 074512 (2020)}.
%[arXiv:1908.06935 [quant-ph]].

\bibitem{ARahman:2021ktn}
S.~A Rahman, R.~Lewis, E.~Mendicelli and S.~Powell,
SU(2) lattice gauge theory on a quantum annealer,
\href{https://doi.org/10.1103/PhysRevD.104.034501}{Phys. Rev. D \textbf{104}, 034501 (2021)}.
%[arXiv:2103.08661 [hep-lat]].

\bibitem{Chandrasekharan:1996ih}
S.~Chandrasekharan and U.~J.~Wiese,
Quantum link models: A Discrete approach to gauge theories,
\href{https://doi.org/10.1016/S0550-3213(97)00006-0}{Nucl. Phys. B \textbf{492}, 455-474 (1997)}.
%[arXiv:hep-lat/9609042 [hep-lat]].

\bibitem{Mathur:2004kr}
M.~Mathur,
Harmonic oscillator prepotentials in SU(2) lattice gauge theory,
\href{https://doi.org/10.1088/0305-4470/38/46/008}{J. Phys. A \textbf{38}, 10015-10026 (2005)}.
%[arXiv:hep-lat/0403029 [hep-lat]].

\bibitem{Byrnes:2005qx}
T.~Byrnes and Y.~Yamamoto,
Simulating lattice gauge theories on a quantum computer,
\href{https://doi.org/10.1103/PhysRevA.73.022328}{Phys. Rev. A \textbf{73}, 022328 (2006)}.
%[arXiv:quant-ph/0510027 [quant-ph]].

\bibitem{Banerjee:2012xg}
D.~Banerjee, M.~B\"ogli, M.~Dalmonte, E.~Rico, P.~Stebler, U.~J.~Wiese and P.~Zoller,
Atomic Quantum Simulation of U(N) and SU(N) Non-Abelian Lattice Gauge Theories,
\href{https://doi.org/10.1103/PhysRevLett.110.125303}{Phys. Rev. Lett. \textbf{110}, 125303 (2013)}.
%[arXiv:1211.2242 [cond-mat.quant-gas]].

\bibitem{Tagliacozzo:2012df}
L.~Tagliacozzo, A.~Celi, P.~Orland and M.~Lewenstein,
Simulations of non-Abelian gauge theories with optical lattices,
\href{https://doi.org/10.1038/ncomms3615}{Nature Commun. \textbf{4}, 2615 (2013)}.
%[arXiv:1211.2704 [cond-mat.quant-gas]].

\bibitem{Zohar:2012xf}
E.~Zohar, J.~I.~Cirac and B.~Reznik,
Cold-Atom Quantum Simulator for SU(2) Yang-Mills Lattice Gauge Theory,
\href{https://doi.org/10.1103/PhysRevLett.110.125304}{Phys. Rev. Lett. \textbf{110}, 125304 (2013)}.
%[arXiv:1211.2241 [quant-ph]].

\bibitem{Zohar:2013zla}
E.~Zohar, J.~I.~Cirac and B.~Reznik,
Quantum simulations of gauge theories with ultracold atoms: local gauge invariance from angular momentum conservation,
\href{https://doi.org/10.1103/PhysRevA.88.023617}{Phys. Rev. A \textbf{88}, 023617 (2013)}.
%[arXiv:1303.5040 [quant-ph]].

\bibitem{Stannigel:2013zka}
K.~Stannigel, P.~Hauke, D.~Marcos, M.~Hafezi, S.~Diehl, M.~Dalmonte and P.~Zoller,
Constrained dynamics via the Zeno effect in quantum simulation: Implementing non-Abelian lattice gauge theories with cold atoms,
\href{https://doi.org/10.1103/PhysRevLett.112.120406}{Phys. Rev. Lett. \textbf{112}, 120406 (2014)}.
%[arXiv:1308.0528 [quant-ph]].

\bibitem{Anishetty:2014tta}
R.~Anishetty and I.~Raychowdhury,
SU(2) lattice gauge theory: Local dynamics on nonintersecting electric flux loops,
\href{https://doi.org/10.1103/PhysRevD.90.114503}{Phys. Rev. D \textbf{90}, 114503 (2014)}.
%[arXiv:1408.6331 [hep-lat]].

\bibitem{Zohar:2014qma}
E.~Zohar and M.~Burrello,
Formulation of lattice gauge theories for quantum simulations,
\href{https://doi.org/10.1103/PhysRevD.91.054506}{Phys. Rev. D \textbf{91}, 054506 (2015)}.
%[arXiv:1409.3085 [quant-ph]].

\bibitem{Mezzacapo:2015bra}
A.~Mezzacapo, E.~Rico, C.~Sab\'\i{}n, I.~L.~Egusquiza, L.~Lamata and E.~Solano,
Non-Abelian $SU(2)$ Lattice Gauge Theories in Superconducting Circuits,
\href{https://doi.org/10.1103/PhysRevLett.115.240502}{Phys. Rev. Lett. \textbf{115}, 240502 (2015)}.
%[arXiv:1505.04720 [quant-ph]].

\bibitem{Silvi:2016cas}
P.~Silvi, E.~Rico, M.~Dalmonte, F.~Tschirsich and S.~Montangero,
Finite-density phase diagram of a (1+1)-d non-abelian lattice gauge theory with tensor networks,
\href{https://doi.org/10.22331/q-2017-04-25-9}{Quantum \textbf{1}, 9 (2017)}.
%[arXiv:1606.05510 [quant-ph]].

\bibitem{Banuls:2017ena}
M.~C.~Ba\~nuls, K.~Cichy, J.~I.~Cirac, K.~Jansen and S.~K\"uhn,
Efficient basis formulation for 1+1 dimensional SU(2) lattice gauge theory: Spectral calculations with matrix product states,
\href{https://doi.org/10.1103/PhysRevX.7.041046}{Phys. Rev. X \textbf{7}, 041046 (2017)}.
%[arXiv:1707.06434 [hep-lat]].

\bibitem{Banerjee:2017tjn}
D.~Banerjee, F.~J.~Jiang, T.~Z.~Olesen, P.~Orland and U.~J.~Wiese,
From the $SU(2)$ quantum link model on the honeycomb lattice to the quantum dimer model on the kagom\'e lattice: Phase transition and fractionalized flux strings,
\href{https://doi.org/10.1103/PhysRevB.97.205108}{Phys. Rev. B \textbf{97}, 205108 (2018)}.
%[arXiv:1712.08300 [cond-mat.str-el]].

\bibitem{Raychowdhury:2018tfj}
I.~Raychowdhury,
Low energy spectrum of SU(2) lattice gauge theory: An alternate proposal via loop formulation,
\href{https://doi.org/10.1140/epjc/s10052-019-6753-0}{Eur. Phys. J. C \textbf{79}, 235 (2019)}.
%[arXiv:1804.01304 [hep-lat]].

\bibitem{Sala:2018dui}
P.~Sala, T.~Shi, S.~K\"uhn, M.~C.~Ba\~nuls, E.~Demler and J.~I.~Cirac,
Variational study of U(1) and SU(2) lattice gauge theories with Gaussian states in 1+1 dimensions,
\href{https://doi.org/10.1103/PhysRevD.98.034505}{Phys. Rev. D \textbf{98}, 034505 (2018)}.
%[arXiv:1805.05190 [hep-lat]].

\bibitem{Raychowdhury:2018osk}
I.~Raychowdhury and J.~R.~Stryker,
Solving Gauss's Law on Digital Quantum Computers with Loop-String-Hadron Digitization,
\href{https://doi.org/10.1103/PhysRevResearch.2.033039}{Phys. Rev. Res. \textbf{2}, 033039 (2020)}.
%[arXiv:1812.07554 [hep-lat]].

\bibitem{Zohar:2019ygc}
E.~Zohar and J.~I.~Cirac,
Removing Staggered Fermionic Matter in $U(N)$ and $SU(N)$ Lattice Gauge Theories,
\href{https://doi.org/10.1103/PhysRevD.99.114511}{Phys. Rev. D \textbf{99}, 114511 (2019)}.
%[arXiv:1905.00652 [quant-ph]].

\bibitem{Raychowdhury:2019iki}
I.~Raychowdhury and J.~R.~Stryker,
Loop, string, and hadron dynamics in SU(2) Hamiltonian lattice gauge theories,
\href{https://doi.org/10.1103/PhysRevD.101.114502}{Phys. Rev. D \textbf{101}, 114502 (2020)}.
%[arXiv:1912.06133 [hep-lat]].

\bibitem{Ji:2020kjk}
Y.~Ji \textit{et al.} [NuQS],
Gluon Field Digitization via Group Space Decimation for Quantum Computers,
\href{https://doi.org/10.1103/PhysRevD.102.114513}{Phys. Rev. D \textbf{102}, 114513 (2020)}.
%[arXiv:2005.14221 [hep-lat]].

\bibitem{Kasper:2020akk}
V.~Kasper, G.~Juzeliunas, M.~Lewenstein, F.~Jendrzejewski and E.~Zohar,
From the Jaynes\textendash{}Cummings model to non-abelian gauge theories: a guided tour for the quantum engineer,
\href{https://doi.org/10.1088/1367-2630/abb961}{New J. Phys. \textbf{22}, 103027 (2020)}.
%[arXiv:2006.01258 [quant-ph]].

\bibitem{Davoudi:2020yln}
Z.~Davoudi, I.~Raychowdhury and A.~Shaw,
Search for efficient formulations for Hamiltonian simulation of non-Abelian lattice gauge theories,
\href{https://doi.org/10.1103/PhysRevD.104.074505}{Phys. Rev. D \textbf{104}, 074505 (2021)}.
%[arXiv:2009.11802 [hep-lat]].

\bibitem{Dasgupta:2020itb}
R.~Dasgupta and I.~Raychowdhury,
Cold-atom quantum simulator for string and hadron dynamics in non-Abelian lattice gauge theory,
\href{https://doi.org/10.1103/PhysRevA.105.023322}{Phys. Rev. A \textbf{105}, 023322 (2022)}.
%[arXiv:2009.13969 [hep-lat]].

\bibitem{Kasper:2020owz}
V.~Kasper, T.~V.~Zache, F.~Jendrzejewski, M.~Lewenstein and E.~Zohar,
Non-Abelian gauge invariance from dynamical decoupling,
\href{https://arXiv.org/abs/2012.08620}{arXiv:2012.08620}.

\bibitem{Ciavarella:2021nmj}
A.~Ciavarella, N.~Klco and M.~J.~Savage,
Trailhead for quantum simulation of SU(3) Yang-Mills lattice gauge theory in the local multiplet basis,
\href{https://doi.org/10.1103/PhysRevD.103.094501}{Phys. Rev. D \textbf{103}, 094501 (2021)}.
%[arXiv:2101.10227 [quant-ph]].

\bibitem{Atas:2021ext}
Y.~Y.~Atas, J.~Zhang, R.~Lewis, A.~Jahanpour, J.~F.~Haase and C.~A.~Muschik,
SU(2) hadrons on a quantum computer via a variational approach,
\href{https://doi.org/10.1038/s41467-021-26825-4}{Nature Commun. \textbf{12}, 6499 (2021)}.
%[arXiv:2102.08920 [quant-ph]].

\bibitem{Kan:2021nyu}
A.~Kan, L.~Funcke, S.~K\"uhn, L.~Dellantonio, J.~Zhang, J.~F.~Haase, C.~A.~Muschik and K.~Jansen,
Investigating a (3+1)D topological \ensuremath{\theta}-term in the Hamiltonian formulation of lattice gauge theories for quantum and classical simulations,
\href{https://doi.org/10.1103/PhysRevD.104.034504}{Phys. Rev. D \textbf{104}, 034504 (2021)}.
%[arXiv:2105.06019 [hep-lat]].

\bibitem{Zohar:2021nyc}
E.~Zohar,
Quantum simulation of lattice gauge theories in more than one space dimension\textemdash{}requirements, challenges and methods,
\href{https://doi.org/10.1098/rsta.2021.0069}{Phil. Trans. A. Math. Phys. Eng. Sci. \textbf{380}, 20210069 (2021)}.
%[arXiv:2106.04609 [quant-ph]].

\bibitem{Raychowdhury:2021jbo}
I.~Raychowdhury,
Toward quantum simulating non-Abelian gauge theories,
\href{https://doi.org/10.1007/s12648-021-02170-6}{Indian J. Phys. \textbf{95}, 1681-1690 (2021)}.
%[arXiv:2106.11475 [hep-lat]].

\bibitem{Klco:2021lap}
N.~Klco, A.~Roggero and M.~J.~Savage,
Standard Model Physics and the Digital Quantum Revolution: Thoughts about the Interface,
\href{https://doi.org/10.1088/1361-6633/ac58a4}{Rept. Prog. Phys. \textbf{85}, 064301 (2022)}.

\bibitem{Kan:2021xfc}
A.~Kan and Y.~Nam,
Lattice Quantum Chromodynamics and Electrodynamics on a Universal Quantum Computer,
\href{https://arXiv.org/abs/2107.12769}{arXiv:2107.12769}.

\bibitem{Ciavarella:2021lel}
A.~N.~Ciavarella and I.~A.~Chernyshev,
Preparation of the SU(3) lattice Yang-Mills vacuum with variational quantum methods,
\href{https://doi.org/10.1103/PhysRevD.105.074504}{Phys. Rev. D \textbf{105}, 074504 (2022)}.
%[arXiv:2112.09083 [quant-ph]].

\bibitem{Illa:2022jqb}
M.~Illa and M.~J.~Savage,
Basic Elements for Simulations of Standard Model Physics with Quantum Annealers: Multigrid and Clock States,
\href{https://arXiv.org/abs/2202.12340}{arXiv:2202.12340}.

\bibitem{Gonzalez-Cuadra:2022hxt}
D.~Gonz\'alez-Cuadra, T.~V.~Zache, J.~Carrasco, B.~Kraus and P.~Zoller,
Hardware efficient quantum simulation of non-abelian gauge theories with qudits on Rydberg platforms,
\href{https://arXiv.org/abs/2203.15541}{arXiv:2203.15541}.

\bibitem{Lewis:2019wfx}
R.~Lewis and R.~M.~Woloshyn,
A qubit model for U(1) lattice gauge theory,
\href{https://arXiv.org/abs/1905.09789}{arXiv:1905.09789}.

\bibitem{Endo:2020kro}
S.~Endo, Z.~Cai, S.~C.~Benjamin and X.~Yuan,
Hybrid Quantum-Classical Algorithms and Quantum Error Mitigation,
\href{https://doi.org/10.7566/JPSJ.90.032001}{J. Phys. Soc. Jap. \textbf{90}, 032001 (2021)}.
%[arXiv:2011.01382 [quant-ph]].

\bibitem{Strikis:2020pcz}
A.~Strikis, D.~Qin, Y.~Chen, S.~C.~Benjamin and Y.~Li,
Learning-based quantum error mitigation,
\href{https://doi.org/10.1103/PRXQuantum.2.040330}{PRX Quantum \textbf{2}, 040330 (2021)}.
%[arXiv:2005.07601 [quant-ph]].

\bibitem{Czarnik:2020xic}
P.~Czarnik, A.~Arrasmith, P.~J.~Coles and L.~Cincio,
Error mitigation with Clifford quantum-circuit data,
\href{https://doi.org/10.22331/q-2021-11-26-592}{Quantum \textbf{5}, 592 (2021)}.
%[arXiv:2005.10189 [quant-ph]].

\bibitem{Lowe}
A.~Lowe, M.~H.~Gordon, P.~Czarnik, A.~Arrasmith, P.~J.~Coles and L.~Cincio,
Unified approach to data-driven quantum error mitigation,
\href{https://doi.org/10.1103/PhysRevResearch.3.033098}{Phys. Rev. Research \textbf{3}, 033098 (2021)}.
%[arXiv:2011.01157 [quant-ph]].

\bibitem{Vovrosh:2021ocf}
J.~Vovrosh, K.~E.~Khosla, S.~Greenaway, C.~Self, M.~Kim and J.~Knolle,
Simple mitigation of global depolarizing errors in quantum simulations,
\href{https://doi.org/10.1103/PhysRevE.104.035309}{Phys. Rev. E \textbf{104}, 035309 (2021)}.
%[arXiv:2101.01690 [quant-ph]].

\bibitem{Urbanek:2021oej}
M.~Urbanek, B.~Nachman, V.~R.~Pascuzzi, A.~He, C.~W.~Bauer and W.~A.~de Jong,
Mitigating depolarizing noise on quantum computers with noise-estimation circuits,
\href{https://doi.org/10.1103/PhysRevLett.127.270502}{Phys. Rev. Lett. \textbf{127}, 270502 (2021)}.
%[arXiv:2103.08591 [quant-ph]].

\bibitem{Sopena:2021sai}
A.~Sopena, M.~H.~Gordon, G.~Sierra and E.~L\'opez,
Simulating quench dynamics on a digital quantum computer with data-driven error mitigation,
\href{https://doi.org/10.1088/2058-9565/ac0e7a}{Quantum Sci. Technol. \textbf{6}, 045003 (2021)}.
%[arXiv:2103.12680 [quant-ph]].

\bibitem{Rosenberg:2021lpj}
E.~Rosenberg, P.~Ginsparg and P.~L.~McMahon,
Experimental error mitigation using linear rescaling for variational quantum eigensolving with up to 20 qubits,
\href{https://doi.org/10.1088/2058-9565/ac3b37}{Quantum Sci. Technol. \textbf{7}, 015024 (2022)}.
%[arXiv:2106.01264 [quant-ph]].

\bibitem{Lloyd}
S.~Lloyd,
Universal quantum simulators,
\href{https://doi.org/10.1126/science.273.5278.1073}{Science {\bf 273}, 1073 (1996)}.

\bibitem{Hatano:2005gh}
N.~Hatano and M.~Suzuki,
Finding Exponential Product Formulas of Higher Orders,
\href{https://doi.org/10.1007/11526216\_2}{Lect. Notes Phys. \textbf{679}, 37 (2005)}.
%[arXiv:math-ph/0506007 [math-ph]].

\bibitem{Wallman}
J.~J.~Wallman and J.~Emerson,
Noise Tailoring for Scalable Quantum Computation via Randomized Compiling,
\href{https://doi.org/10.1103/PhysRevA.94.052325}{Phys. Rev. A \textbf{94} 052325 (2016)}.

\bibitem{Bravyi}
S.~Bravyi, S.~Sheldon, A.~Kandala, D.~C.~Mckay, and J.~M.~Gambetta,
Mitigating measurement errors in multiqubit experiments,
\href{https://doi.org/10.1103/PhysRevA.103.042605}{Phys. Rev. A \textbf{103} 042605 (2021)}.

\bibitem{qiskit}
M.~S.~Anis et al.,
{\em Qiskit: An Open-source Framework for Quantum Computing},
\href{https://doi.org/10.5281/zenodo.2573505}{doi:10.5281/zenodo.2573505 (2021)}.

\bibitem{Li}
Y.~Li and S.C.~Benjamin,
Efficient Variational Quantum Simulator Incorporating Active Error Minimization,
\href{https://doi.org/10.1103/PhysRevX.7.021050}{Phys. Rev X \textbf{7} 021050 (2017)}.

\bibitem{duet} \href{https://github.com/randylewis/SelfMitigation}{https://github.com/randylewis/SelfMitigation}

\end{thebibliography}
\end{document}